\documentclass[%
 reprint,
 amsmath,amssymb,
 aps,
prd,
]{revtex4-2}

\usepackage{graphicx}
\usepackage{dcolumn}
\usepackage{bm}
\usepackage[dvipsnames]{xcolor}
\usepackage[unicode]{hyperref}
\hypersetup{colorlinks=true, citecolor=MidnightBlue,
            linkcolor=MidnightBlue, urlcolor=MidnightBlue, linktocpage=true}
\usepackage{multirow}
\usepackage{array}
\usepackage{enumitem}
\usepackage{mathtools}

\begin{document}

\title{Overcharging Extremal Rotating Black Holes}
\author{Shauvik Biswas\footnote{shauvikbiswas2014@gmail.com}$~^{1}$, Amruta Sadhu\footnote{amruta.sadhu@xaviers.edu}$~^{2}$ and Sudipta Sarkar\footnote{sudiptas@iitgn.ac.in}$~^{1}$
\\
$^{1}${\small{Indian Institute of Technology, Gandhinagar, Gujarat 382055, India}}\\
$^{2}${\small{St. Xavier's College, Mumbai, Maharashtra, 400001, India}}}
\email{shauvikbiswas2014@gmail.com}
\email{amruta.sadhu@xaviers.edu}
\email{sudiptas@iitgn.ac.in}

\begin{abstract}
In this work, we use the weak cosmic censorship conjecture (WCC) to constrain black hole solutions in modified gravity theories. While Wald showed that extremal Kerr–Newman black holes in general relativity cannot be overcharged by test charged particles, this protection may fail in theories beyond general relativity. We have considered generic rotating black hole solutions beyond the Kerr–Newman family and examined particle absorption processes that could lead to overcharging and the emergence of naked singularities. Identifying regions of parameter space where WCC is violated allows us to place direct, physically motivated bounds on deviations from general relativity.
\end{abstract}

\maketitle

\section{Introduction}
General Relativity (GR) stands as the most successful theory of gravity, describing spacetime as a Lorentzian manifold \cite{Hawking:1973uf}. However, it faces several fundamental challenges, including the existence of singularities \cite{PhysRevLett.14.57}, non-renormalizability \cite{Ramond:1981pw}, and the apparent incompatibility between the equivalence principle and quantum unitarity, as suggested by the quantum evaporation of black holes \cite{PhysRevD.14.2460}. These issues indicate that GR could be an effective theory \cite{Donoghue:1994dn}, analogous to how the Fermi theory of beta decay serves as an effective description of weak interactions, hinting at the necessity of a more fundamental theory of gravity.\\

Any modification to GR generally introduces terms involving powers of curvature tensors. However, such higher-curvature terms need not be arbitrary. For instance, one could impose the condition that only those terms are allowed that yield equations of motion that remain second order in time derivatives \cite{PhysRevD.91.085009,Woodard:2015zca}.
In this context, Lovelock theorem \cite{Lovelock:1971yv} states that in four spacetime dimensions, only second-ranked symmetric tensors that are made from metric and its derivative and are divergence free in one index are the metric and the Einstein tensor. This criterion, together with the equivalence principle, uniquely selects GR as the only viable theory in $3+1$ dimensions. Alternatively, within the framework of effective field theory, one would systematically include all possible higher-curvature terms and analyse their effects perturbatively, order by order.\\

Observationally, some constraints on such extensions of GR already exist. LIGO-VIRGO observations provide a weak constraint on the higher-curvature coupling constant, for example, in the case of Einstein-dilaton-Gauss-Bonnet(EdGB) theory the coupling constant, $\alpha_{\rm GB}$ is constrained by $\sqrt{\alpha_{GB}} < 1.7$ km \cite{Perkins:2021mhb}. Additional constraints arise from black hole shadow observations\cite{Ghosh:2022jfi}, local tests of gravity \cite{PhysRevD.102.084005,Bellucci:2018yuz,Odintsov:2022zrj}, and cosmological probes \cite{Zanoletti:2023ori,Neupane:2007qw}. However, all such constraints remain rather weak at present. Therefore, alternative approaches are necessary to impose stricter constraints on modifications of gravity. Notably, this can be accomplished by enforcing theoretical consistency conditions and ensuring the well-posedness of the theory's dynamics. For instance, the validity of the black hole second law could serve as a constraint on modified gravity theories \cite{Sarkar:2019xfd}. Additionally, the sign of the Gauss-Bonnet coupling can be determined by requiring unitarity in the spectral representation \cite{Cheung:2016wjt,Caron-Huot:2022ugt}.  \\

In this study, we explore an important additional consistency condition, namely the validity of the weak cosmic censorship (WCC) conjecture \cite{Hawking:1994ss} as a means to assess the self-consistency of gravity theories beyond general relativity. The WCC conjecture asserts that singularities capable of causally influencing the asymptotic regions of spacetime must be concealed. Specifically, during gravitational collapse, a naked singularity should not form, and all singularities must remain hidden behind a black hole horizon.\\

Since a general proof of this conjecture is lacking \cite{Wald:1997wa,Joshi:2002dt}, research often focuses on analyzing potential counterexamples. The typical approach involves starting with an extremal or near-extremal black hole and examining whether any physical process could lead to the formation of a naked singularity. If such a process is possible, it would indicate a potential violation of the WCC conjecture\cite{Yang:2025leh}. The first attempt to test this was made in a foundational study by Wald, who demonstrated that it is impossible to overcharge or overspin an extremal Kerr-Newman black hole through the absorption of test particles \cite{Wald:1974hkz}. Subsequent research has extensively examined the validity of the WCC conjecture for various classes of black hole spacetimes \cite{Hod:2008zza,Fairoos:2017lnm,Jacobson:2009kt}. One notable study by Hubeny suggested that overcharging a near-extremal black hole to create a naked singularity is possible under the assumption that back-reaction effects are negligible \cite{Hubeny:1998ga}. However, a more recent analysis addressed this overcharging problem in detail. Assuming the null energy condition on matter, it was shown that forming a naked singularity by overcharging extremal or near-extremal black holes becomes impossible when back-reaction effects are considered \cite{Sorce:2017dst}.\\

Despite extensive investigations, most studies have concentrated on black hole solutions within general relativity, leaving the status of the weak cosmic censorship (WCC) conjecture in alternative gravity theories largely unexplored. A notable exception is the case of charged black holes in Einstein-Gauss-Bonnet gravity, where the conjecture has been explicitly examined \cite{Ghosh:2019dzq}. This result has also been extended to all Lovelock gravity theories \cite{Shaymatov:2020byu}. Interestingly, analyses of the WCC conjecture for a spherically symmetric general black hole solution have revealed constraints on the parameters characterizing deviations from general relativity \cite{Ghosh:2021cub}. Such deviations may arise due to the presence of higher-curvature terms or as post-Einsteinian corrections to general relativity.

In this work, we provide a significant generalization of these previous results (mainly as in \cite{Ghosh:2021cub}) by considering a general rotating solution from a phenomenological perspective, ensuring its correspondence with the Kerr-Newman metric. We then derive possible constraints on the metric components that prevent such an extremal black hole solution from being overcharged by a test body. This, in turn, leads to intriguing restrictions on modified gravity models that admit such black hole solutions.

This paper is organized as follows: In section.\ref{Sec-1} we will propose the theory agnostic version of rotating black hole spacetime metric. Subsequently, in section.\ref{sec-2} we will demonstrate overcharging and entering conditions of such an extremal rotating black hole by test charged particles. Then, in section.\ref{sec-3} we will study the status of WCC in such spacetime by considering only the leading order correction term. Finally, we have discussed our key findings. \\

Throughout the paper, we will consider the metric signature to be mostly positive and set $c=G=1$.


\section{Proposal for Theory-Agnostic Rotating Metric}
\label{Sec-1}

In realistic astrophysical situations, the collapse of massive stars typically results in rotating black holes due to non-vanishing net angular momentum. Within vacuum general relativity, the Kerr solution provides the unique, stationary, axisymmetric, and asymptotically flat black hole metric. However, in modified theories of gravity, such uniqueness theorems generally do not exist. Consequently, one could adopt a correspondence principle: any viable rotating solution in the modified theory should reduce to the Kerr metric in the appropriate limit, i.e., when the coupling constants associated with the deviations vanish. In this spirit, given the observational success of general relativity, the deviations from Kerr are expected to appear perturbatively, order by order in the new coupling constants.

To construct such rotating solutions, a common approach is to begin with a static, spherically symmetric black hole solution in the modified theory. Then one attempts to apply the Newman-Janis algorithm (NJA) to generate its rotating counterpart. Unfortunately, this method often encounters obstacles: for example, the coordinate transformation from advanced (or retarded) Eddington–Finkelstein time to Boyer–Lindquist time may not exist, or the resultant rotating metric may fail to satisfy the modified field equations (see \cite{Biswas:2022wah} for a mathematical exposition of these limitations). Furthermore, the NJA may yield non-unique solutions since the algorithm depends on the specific complexification prescription for the radial coordinate.

Recent works \cite{Azreg-Ainou:2014aqa, PhysRevD.90.064041, Junior:2020lya} propose modified versions of the NJA, leading to more general classes of axisymmetric metrics. Motivated by these developments, we consider the following most general ansatz for an axisymmetric black hole metric with mass $M$, charge $Q$, and angular momentum $J$:
\begin{align}
\label{gen-metric}
ds^2 &= -F\, dt^2 - 2a \sin^2\theta \left(\sqrt{\frac{F}{G}} - F\right) dt\, d\phi \nonumber \\
&\quad + H\, d\theta^2 + \frac{H}{G H + a^2 \sin^2\theta}\, dr^2 \nonumber \\
&\quad + \sin^2\theta \left[H + a^2 \sin^2\theta \left(2\sqrt{\frac{F}{G}} - F\right)\right] d\phi^2,
\end{align}
where the functions $F$, $G$, and $H$ depend on the coordinates $(r,\theta)$. The rotational parameter $a = J/M$ characterizes the specific angular momentum of the black hole. Notably, the above metric represents a special subclass of the general family proposed in \cite{PhysRevD.90.064041}.\\

In general, for a stationary spacetime, the \textit{strong rigidity theorem} need not be valid, and as a result, the \textit{global event horizon} does not necessarily coincide with the \textit{Killing horizon}. The proof of the rigidity theorem relies fundamentally on the Einstein field equations, and therefore, once these equations are modified or replaced, as in theories beyond general relativity, the theorem’s applicability becomes uncertain. In the absence of such a result, even the precise formulation of the \textit{cosmic censorship conjecture} becomes somewhat ambiguous, since the causal structure of the horizon may no longer be well defined. \\

Nevertheless, there are indications that suitably generalized versions of these theorems might survive beyond general relativity, particularly within the framework of \textit{effective field theories} (EFT) of gravity~\cite{Hollands:2022ajj}. In fact, analyses within the same EFT framework suggest that all stationary black hole solutions in any metric theory of gravity must possess \textit{circular symmetry}, which, in \textit{Boyer--Lindquist coordinates}, corresponds to invariance under the simultaneous transformation $(t, \phi) \rightarrow (-t, -\phi)$ \cite{Xie:2021bur}. From the observational perspective also, gravitational wave detections have placed stringent constraints on any possible deviations from the Kerr geometry \cite{Ghosh:2024het}. Taken together, these theoretical considerations and empirical restrictions provide strong motivation for our \textit{metric ansatz}, which respects the essential symmetry and structure of stationary, axisymmetric black holes while allowing controlled deformations to test possible departures from general relativity.

Motivated by this, we demand that the event horizon of the metric \eqref{gen-metric} also serves as a Killing horizon. This condition implies that the norm of the Killing vector field $\xi^\mu = \xi^\mu_{(t)} + \Omega_H \xi^\mu_{(\phi)}$ must vanish on the horizon, where $\xi^\mu_{(t)} = \left(\partial_t\right)^\mu$ and $\xi^\mu_{(\phi)} = \left(\partial_\phi\right)^\mu$ are the time-translation and rotational Killing vectors, respectively. The angular velocity of the horizon is given by
\begin{align}
\Omega_H = -\left.\frac{g_{t\phi}}{g_{\phi\phi}}\right|_{\mathcal{H^+}}.
\end{align}
The assumption of rigidity ensures that the angular velocity $\Omega_H$ is a constant on the horizon.\\

We would like to emphasise that our results are entirely independent of the validity or applicability of the Newman–Janis algorithm (NJA). The spacetime employed in our analysis is already well-motivated and formulated in a sufficiently general framework. In particular, our metric encompasses a broad class of stationary and axisymmetric black hole geometries that are consistent with the minimal physical assumptions of asymptotic flatness, horizon regularity, and rigidity. Hence, the conclusions we draw do not rely in any way on the correctness of the NJA procedure, but rather on the intrinsic geometric and physical generality of the spacetime we consider.\\

This leads to the definition of a horizon function, $f(r,\theta;M,a,Q) \equiv G H + a^2 \sin^2\theta$, whose vanishing determines the location of the event horizon. In the extremal limit, the function $f(r,\theta;M,a,Q) = 0$ must admit a unique root for each value of $\theta$, leading to the additional requirement: $\partial_r f(r,\theta;M,a,Q) = 0$ on the horizon. These two conditions must be solved simultaneously to determine the location of the extremal event horizon, which in general may depend on the polar angle $\theta$.

In the remainder of this work, we consider the metric Eq.\eqref{gen-metric} as the foundational geometry for deriving criteria for particle infall and overcharging. We will analyze these conditions in the most general form and then apply them to specific higher curvature theories. 


\section{Basic setup: overcharging and entering conditions}\label{sec-2}
In this section, we are interested in performing a thought experiment, that is, we want to analyze if we can overcharge and over spin the black hole by throwing test particles into the black hole along the axis of rotation. To begin with, we consider an extremal rotating black hole with hairs $(M,a,Q)$ such that the above stated extremality condition holds for all values of $\theta$ within the range $0\leq\theta\leq\pi$. Now we throw a test particle of energy $\delta E$ satisfying $0<\delta E << M$, charge $\delta Q$ satisfying $0<\delta Q<<Q$ and angular momentum $\delta J$ satisfying $0<\delta J<<J$ into the black hole. To the leading order in the energy and angular momentum of the particle, the increment of the rotation parameter of the black hole is,
\begin{align}\label{increment-a}
    \delta a=\frac{\delta J}{M}-a\frac{\delta E}{M}~.
\end{align}
If $u(\theta,a,M,Q)$ denotes the horizon radius of the extremal rotating black hole, then the final object will be a naked singularity if $f(u+\delta u,\theta,M+\delta E, Q+\delta Q,a+\delta a )>0$. If this condition holds then the black hole can be overcharged/overspun.~Now to leading order in $\delta E$, $\delta Q$ and $\delta a$ the above inequality reduces to,
\begin{align}\label{Overcharging-2}
    \delta E\left(\partial_{M}f(u,\theta)-\frac{a}{M}\partial_{a}f(u,\theta)\right)&+\delta Q\,\partial_{Q}f(u,\theta)\\
    \nonumber
    +&\frac{\delta J}{M}\partial_{a}f(u,\theta) > 0~.
\end{align}
This is the so-called overcharging and/or overspinning condition.~In deriving this condition we have exploited the extremality condition stated in the previous section and used Eq.(\ref{increment-a}). Specifying this condition does not complete all the requirements and we must also ensure that such a particle be captured by the black hole. This requirement can be obtained by the kinematic condition that there should be no turning point on the trajectory of the test particle till it reaches the horizon. This leads to the condition \cite{Wald:1974hkz},
\begin{align}\label{Entering-condition}
    \delta E \geq -\delta Q A_{t} + \frac{g^{t\phi}}{g^{tt}}(\delta J -\delta Q A_{\phi})~.
\end{align}
It is the so-called entering condition.~Thus conditions Eq.(\ref{Entering-condition}) and Eq.(\ref{Overcharging-2}) must be satisfied simultaneously to overcharge the black hole and form a naked singularity.\\

Now depending on the choice of signs of $[\partial_{M}f,\, \partial_{a}f,\, \partial_{Q}f]$, we can have eight different cases for the overcharging condition Eq.(\ref{Overcharging-2}). Interestingly, the entering condition Eq.(\ref{Entering-condition}) always imposes lower bound on $\delta E$. Thus, to obtain non-trivial parameter space for overcharging, we must choose those sign combinations for which the overcharging condition imposes, upper bound on $\delta E$. The remaining combinations will be in conflict with the entering condition and never lead to overcharging/overspinning. In a straightforward calculation, we calculate and tabulate in the appendix~\ref{sec:appA}, all such non-trivial cases for which we have \textit{positive} upper bound on $\delta E$.\\

The general strategy, therefore, is to begin with a specific black hole solution within a given modified theory of gravity. One must first determine the appropriate sign combination of the derivatives $[\partial_{M}f,\, \partial_{a}f,\, \partial_{Q}f]$, and subsequently assess whether overcharging is possible. If, in contrast to general relativity, an extremal black hole can be overcharged, resulting in the formation of a naked singularity and thereby violating the Weak Cosmic Censorship (WCC) conjecture, then such a theory is deemed untenable. Moreover, this analysis can impose significant constraints on the allowed parameter space of the underlying modified theory.\\

In the next section, we will consider special metric functions which can be motivated from modified theories of gravity and use the above stated conditions such that the modified theories obey weak cosmic censorship.

\section{Testing WCC in modified rotating black hole spacetime}\label{sec-3}

As discussed earlier, the uniqueness theorems in general relativity assert that vacuum, stationary, and axisymmetric black hole solutions are fully characterized by only three parameters: mass, charge, and angular momentum. However, in general modified theories of gravity, such uniqueness theorems are absent, leading to a significantly enlarged solution space. Despite this, no exact rotating black hole solutions are known in any modified gravity theory beyond the Kerr family. While some approximate solutions to the linearized field equations exist, fully non-linear, exact solutions remain elusive. Consequently, our analysis relies on a physically well-motivated ansatz for the metric functions that extends beyond the standard Kerr-Newman class. Furthermore, we require that these solutions asymptotically reduce to the Kerr-Newman geometry to ensure consistency with general relativity in the appropriate limit.\\

In this paper, we will consider black hole solutions of the modified theories in 3+1 dimensions which have the following metric functions,
\begin{align}\label{Metric-functions}
 &H(r,\theta)=\rho^2=r^2+a^2 \cos^2{\theta}~,\\
 & F(r,\theta)=G(r,\theta)= \\
  &1-\frac{2 M r}{\rho^{2}}+\frac{Q^2}{\rho^{2}}+\sum_{k=2}^{\infty}\left[\alpha_{2k-1} \frac{r}{M}+\alpha_{2k}\right]\left(\frac{M^2}{\rho^2}\right)^{k}\nonumber~.
\end{align}
Here, $M$, $Q$ and $a$ have usual interpretations and $\alpha_i$s' denotes the additional parameters which may appear in modified gravity models. As a concrete example, consider \textit{slowly rotating} black holes in the Einstein-\AE{}ther theory \cite{Zhu:2019ura}, which is given in the form, 
\begin{flalign*}
g_{tt} &= -\frac{1}{g_{rr}} = -\left(1 - \frac{2M}{r} + \frac{Q}{r^{2}} + \alpha_{4}\left(\frac{2M}{r}\right)^{4}\right) &&\\
&g_{t\phi} = -\frac{2M}{r}a\sin^{2}\theta &&\\
&g_{\theta\theta} = r^{2}, \,\, g_{\phi\phi} = r^{2}\sin^{2}\theta~. &&
\end{flalign*}
Where, the constant $\alpha_{4}$ is related to the various coupling constants of Einstein-\AE{}ther theory. Using Eq.[\ref{Metric-functions}] and taking slowly rotating limit, one can easily check that the above metric components are of the form given in Eq.[\ref{gen-metric}].
Such a metric was also proposed in the context of Kerr-like solutions \cite{PhysRevD.83.124015} as well as in case of Post-Newtonian corrections \cite{Will:2014kxa}. We emphasize that, we can not directly motivate this modified gravity black hole metric using Newman-Janis algorithm due to previously stated pathologies, however one can easily notice that the metric reduces to Kerr metric as $\alpha$s' goes to zero. Moreover, one can check that this spacetime has regular horizons, that is curvature scalars remain finite on the horizon. Now let us quote the components of the vector potential $A_{\mu}$.  Since we want our spacetime to have a continuous correspondence with the symmetries of Kerr-Newman spacetime, the vector potential is of the same form as that of the Kerr-Newman metric, that is,
\begin{align}
    A_{\mu}=\left(-\frac{Qr}{\rho^{2}},0,0,\frac{Qra \sin^{2}\theta}{\rho^{2}}\right)~.
\end{align}
Therefore, the non-vanishing components of the electromagnetic tensor will be 
\begin{align}
    &F_{rt}=\frac{Q}{\rho^{4}}(r^{2}-a^{2}\cos^{2}\theta)~,F_{r\phi}=\frac{Q a \sin^{2}\theta}{\rho^{4}}(r^{2}-a^{2}\cos^{2}\theta)~,\\
    &F_{\theta t}=-\frac{2Qa^{2}r}{\rho^{4}}\sin\theta\cos\theta~,\\
    &F_{\theta\phi}=\frac{Qra(r^{2}+a^{2})\sin(2\theta)}{\rho^{4}}~.
\end{align}

It is important to note that the assumption of asymptotic flatness enables the definition of the total mass \( M \) and angular momentum \( J \) through the standard Arnowitt-Deser-Misner (ADM) formalism which should remain applicable in any reasonable modified theory of gravity. Furthermore, the total electric charge \( Q \) can be defined via Gauss’s law, which relies solely on the behavior of the electromagnetic field at spatial infinity and is thus independent of the underlying gravitational theory. \\

Now, we start with an extremal black hole of this class, and aim to overcharge it using a test charged particle \footnote{Here, for simplicity, we assume that the text particle does not impart any angular momentum, i.e. $\delta J = 0$.}. Our aim would be to constrain the extra parameters $\alpha_i$s using the requirement of the validity of WCC. To show how this can be achieved, let us first consider the simplest deviation from the Kerr metric by keeping only $\alpha_{3}$ as deviation parameter. In this case, upto a $r^{2}$ factor the horizon function reads,
\begin{align}\label{Our-Horizon}
f(r,\theta,M,a,Q,\alpha_{3})=1-\frac{2M}{r}+\frac{a^2+Q^2}{r^2}+\frac{\alpha_3 M^3}{r\rho^2}~.
\end{align}
 From this expression, it is clear that $\partial_{Q}f > 0$, which implies that we have to consider the cases with sign combinations for $[\partial_{M}f, \partial_{a}f, \partial_{Q}f]$ as $[-,+,+]$, $[+,+,+]$ and $[-,-,+]$ and analyze the status of WCC. \\
 
 At this position, it is worth to note that the extremality condition $\partial_{r}f(u)=0$ leads to 
 \begin{align}\label{extremality}
 \alpha_{3}M^{3}=\frac{2\rho_{u}^{4}}{\left(2u^{2}+\rho^{2}\right)}\left[M-\frac{a^{2}+Q^{2}}{u}\right]~,
 \end{align}
 where $u$ stratifies $f(u)=0$ for all $\theta$ and $\rho^{2}=u^{2}+a^{2}\cos^{2}\theta$. Then it follows that for $\alpha_{3}>0$, we get the condition $M u > (a^{2}+Q^{2})$ and the opposite inequality holds for $\alpha_{3}<0$. On the other hand, for $\alpha_{3}=0$ we get $Mu=(a^{2}+Q^{2})$ which with the condition $f(u)=0$ leads to the known result $u=M$ for general relativity.
 \\
 
 Let us first consider the case $[-,+,+]$, which also occurs in the Kerr-Newmann case also. Consider, that a particle having no angular momentum is falling along the axis of rotation. Then, from Eq. (\ref{-++}), we have the overcharging condition:
 \begin{align}\label{Overcharging-Our-Condition}
  \delta E <\frac{\delta Q \partial_{Q}f}{|\partial_{M}f|+\frac{a}{M}\partial_{a}f}~.
 \end{align}
The test particles will follow the geodesics of the background modified metric. However, the test charged particle can sense the background electromagnetic field, so that they will follow the below equation of motion[40],
\begin{align}\label{EOM-Particle}
    u^{\mu}\nabla_{\mu}u^{\nu}=\frac{\delta Q}{M}F^{\nu\alpha}u_{\alpha}~.
\end{align}
Here $\delta Q$ is the particle's charge and $u^{\mu}$ is the particle's four velocity. Also, the covariant derivative is compatible with the background modified metric.

 To analyze the entering condition Eq. (\ref{Entering-condition}), let us also consider the same case, when the particle is falling along the z-axis, i..e, $\theta = 0$, which gives the greatest lower bound. Therefore, using the expressions of the vector potentials $A_{t}$ and $A_{\phi}$, we obtain the entering condition as from Eq.[\ref{EOM-Particle}], 
 \begin{align}\label{Entering-our-case}
 \delta E> \frac{Qu }{u^{2}+a^{2}}\delta Q~.
 \end{align}
We emphasize that this special condition is chosen to obtain the greatest bound on the parameters, so that the constraints Eq.(\ref{Entering-condition}) hold for all values of theta.  This can be easily seen if we consider that on the black hole horizon,
\begin{align}
    -A_{t}-\frac{g^{t\phi}}{g^{tt}}A_{\phi}=\frac{Q\, u}{\rho_{u}^{2}}\left[1-\frac{a^{2}\sin^{2}\theta(u^{2}+a^{2})}{\rho_{u}^{2}(u^{2}+a^{2})+a^{2}\sin^{2}\theta(u^{2}+a^{2})}\right]
\end{align}
which maximizes for $\theta=0$ and yields the above bound.\\

Next, from Eq. (\ref{Overcharging-Our-Condition}) and Eq. (\ref{Entering-our-case}), it follows that the black hole can be overcharged if 
 \begin{align}\label{Main-ineqality-our}
  a^{2}(u-M)<u\,(Mu-a^2-Q^2)\left( 1+ \frac{4 a^{2}\cos^2\theta}{2u^2+\rho^{2}_{u}}\right)~.
 \end{align}
 Now this inequality is not independent of the extremality condition 
 Eq. (\ref{extremality}), that is the above condition also must hold for those values of $u$ which satisfies the extremality condition Eq. (\ref{extremality}). This opens up the possibility of the consideration of two cases, namely $\alpha_{3}>0$ and $\alpha_{3}<0$. \\
 
  \emph{\textbf{Case-1 ($\alpha_{3}>0$)}:} First, let us consider the case $u<M$, so that $(u-M)<0$ and from Eq. (\ref{extremality}), we have $Mu>(a^{2}+Q^{2})$. Thus it follows that the inequality Eq. (\ref{Main-ineqality-our}) holds and the black hole can be overcharged. Opposite case is a bit more intricate. In this case, the horizon equation $f(u)=0$ and the extremality condition Eq. (\ref{extremality}) leads to the condition,
  \begin{align}
  u^{2}(u-M)>a^{2}(u-M)~.
  \end{align}
  On the other hand due to $\alpha_{3}>0$, the horizon condition yields,
  \begin{align}
  (Mu-a^{2}-Q^{2})\, u> u^{2}(u-M)~.
  \end{align}
  Also, since $0\leq \cos^{2}\theta\leq 1$, we have,
  \begin{align}
  u(Mu-a^2-Q^2)\left( 1+ \frac{4 a^{2}\cos^2\theta}{2u^2+\rho^{2}_{u}}\right)\geq  u(Mu-a^2-Q^2)~.
  \end{align}
Combining the above three inequalities one can show that Eq. (\ref{Main-ineqality-our}) can be satisfied, that is the black hole can be overcharged if $\alpha_{3}>0$. \\

\emph{\textbf{Case-2 ($\alpha_{3}<0$)}:} In this case, the extremality condition dictates that, $Mu-(a^{2}+Q^{2})<0$. To proceed further, we rewrite the 
Eq. (\ref{Main-ineqality-our}) as,
\begin{align}\label{Main-ieq-Modified}
(a^{2}+Q^{2}-Mu)&\frac{4 \, u\, a^{2}\cos^{2}\theta}{2 u^{2}+\rho^{2}_{u}}\nonumber\\
&< a^{2}(M-u)-(a^{2}+Q^{2}-Mu)u~.
\end{align}
Now for $u>M$, the right hand side is negative and the left hand side positive, leading to a contradiction. Thus for $\alpha_{3}<0$ the black hole can not be overcharged if $u>M$. To explore the situation for the case $u<M$, we first algebraically combine the horizon equation and the exremality condition to show that,
\begin{align}
 \alpha_{3} M^3=\frac{2(u-M)\rho_u^4}{(u^2-a^2\cos^2\theta)}~,
\end{align} 
from which it follows that $u^{2}>a^{2}$. Now the horizon equation dictates that $ (Mu-a^{2}-Q^{2})u< u^{2}(u-M)$. Using these two inequalities with the extremality condition we can show that,
\begin{align}
a^2(M-u)-(a^2+Q^2-Mu)u<\frac{(a^2+Q^2-Mu)}{u}(a^2-u^2).
\end{align}
Thus for overcharging to be possible, the left hand side of Eq. (\ref{Main-ieq-Modified}) must at least satisfy,
\begin{align}
(a^{2}+Q^{2}-Mu)\frac{4 u a^{2}\cos^{2}\theta}{2 u^{2}+\rho^{2}_{u}}<\frac{(a^2+Q^2-Mu)}{u}(a^2-u^2)~.
\end{align}
This immediately leads to the contradiction for $u^{2}>a^{2}$. Thus the black hole can not be overcharged for $\alpha_{3}<0$.\\

Now we are left with two more cases, namely $[+,+,+]$ and $[-,-,+]$. But for the case $[+,+,+]$, we have an additional restriction given by the condition, $\partial_{M}f(u)<\frac{a}{M}\partial_{a}f(u)$. Using it, the denominator of 
Eq. (\ref{+++}) becomes,
\begin{align}
|\partial_{M}f-\frac{a}{M}\partial_{a}f|=\frac{2}{u}+\frac{2 a^2}{M u^2}-\frac{\alpha_{3}M^{2}}{u\rho_{u}^{2}}\left(3+\frac{2a^{2}\cos^{2}\theta}{\rho_{u}^{2}}\right)~.
\end{align}
Similarly, for the case $[-,-,+]$, we have the denominator,

\begin{align}
|-|\partial_{M}f|+\frac{a}{M}|\partial_{a}f||=&|\partial_{M}f|-\frac{a}{M}|\partial_{a}f|=\\
&-\partial_{M}f+\frac{a}{M}\partial_{a}f\nonumber~.
\end{align}

This expression is identical to the one obtained previously. Consequently, the same line of reasoning used to test the validity of the Weak Cosmic Censorship (WCC) conjecture applies to the remaining two cases as well. Specifically, black holes cannot be overcharged when $\alpha_3 < 0$, implying that WCC remains intact in such scenarios. Therefore, modified gravity theories predicting $\alpha_3 < 0$ are consistent with cosmic censorship and hence considered viable, as in the case of general relativity. Conversely, theories with $\alpha_3 > 0$, which allow overcharging and hence a potential violation of WCC, can be deemed theoretically disfavored.\\

This illustration serves as a concrete example of how the validity of the Weak Cosmic Censorship (WCC) conjecture can impose significant constraints on the admissible structure of black hole solutions in theories extending beyond general relativity. Owing to the generality of our formulation, the same line of analysis can be straightforwardly applied to a broad class of such black hole spacetimes, enabling systematic tests of WCC in diverse modified gravity scenarios.

\section{Discussions}\label{4}

Despite its remarkable success in describing gravitational phenomena in the weak-field regime, the validity of general relativity (GR) in the strong-field domain remains an open question, particularly where quantum gravitational effects are expected to become significant. In this context, GR may be viewed as an effective low-energy theory, analogous to Fermi's theory of beta decay. Modifications to the Einstein–Hilbert action that aim to provide a more accurate low-energy description typically involve higher-curvature corrections or the introduction of additional fields, as seen in scalar–tensor or vector–tensor theories. However, current observational capabilities are insufficient to fully constrain the vast landscape of such higher-order couplings and alternative theories.\\

In this work, we employ the Weak Cosmic Censorship (WCC) conjecture as a theoretical criterion to constrain such modified theories. The WCC posits that gravitational collapse should not lead to the formation of naked singularities; instead, singularities must be hidden behind event horizons, preserving the causal structure of spacetime. A seminal result by Wald \cite{Wald:1974hkz} demonstrated that extremal Kerr–Newman black holes cannot be overcharged or overspun by test particles, thus upholding WCC within classical GR. However, recent studies \cite{Ghosh:2021cub} indicate that in modified black hole spacetimes, where additional `hair' arises from higher-curvature terms or extra dynamical fields, there may exist non-trivial regions in parameter space that permit overcharging, thereby violating WCC. Such violations can then be used to place theoretical constraints on the viability of the corresponding modified theories.\\

 In the present work, we construct a class of modified rotating, asymptotically flat black hole solutions that retain regular horizons and reduce to the Kerr–Newman geometry in the limit of vanishing deviations. We specifically consider extremal configurations and incorporate leading-order corrections to the Kerr–Newman metric. Our analysis reveals that for certain sign combinations of these correction terms, both the entering and overcharging conditions can be simultaneously satisfied, thus enabling a potential violation of WCC. This approach allows us to systematically constrain the parameter space of the modified black hole solutions and, by extension, the underlying modified gravity theories.\\

A natural avenue for future research is to incorporate the effects of backreaction and self-force, both of which are expected to play a pivotal role in any realistic scenario involving infalling matter. While the test particle approximation offers valuable insight into potential violations of the Weak Cosmic Censorship (WCC) conjecture, it is well established that including the particle’s backreaction on the spacetime geometry can, in many instances, restore cosmic censorship. Consequently, a perturbative analysis that accounts for second-order effects, or fully nonlinear numerical relativity simulations adapted to modified gravity frameworks, will be essential for evaluating the robustness of our present conclusions. Furthermore, beyond general relativity, there may exist black hole solutions that do not possess axisymmetry. Extending our analysis to such non-axisymmetric black hole spacetimes would provide a broader testing ground for WCC. Taken together, these extensions would contribute to a more comprehensive understanding of how fundamental theoretical consistency conditions constrain the space of viable modifications to general relativity.

\section*{Acknowledgement}
We thank Rajes Ghosh for helpful discussions.  S.S. extends his gratitude to the Institute of Theoretical Physics (ITP), Chinese Academy of Sciences for their excellent hospitality, where a part of this work was completed. The research of S.S. is partially funded by the Department of Science and Technology, Government of India, through the SERB CRG Grant (No. CRG/2023/000545). 
\appendix

\section{}\label{sec:appA}
Here we tabulate the choice of signs of $[\partial_{M}f,\partial_{a}f,\partial_{Q}f]$, such that there is a possibility of satisfying the entering condition. The allowed choices are:

\begin{enumerate}[label=\arabic*.]
    \item $\left[\partial_{M}f,\,\partial_{a}f,\,\partial_{Q}f\right] \equiv [+,+,+]$:
    \begin{equation}\label{+++}
        \delta E < \frac{ \delta Q\, \partial_Q f + \dfrac{\delta J}{M} \partial_a f }
                        { \left| \partial_M f - \dfrac{a}{M} \partial_a f \right| }, 
        \quad \text{with } \partial_M f < \dfrac{a}{M} \partial_a f~.
    \end{equation}

    \item $\left[\partial_{M}f,\,\partial_{a}f,\,\partial_{Q}f\right] \equiv [-,+,+]$:
    \begin{equation}\label{-++}
        \delta E < \frac{ \delta Q\, \partial_Q f + \dfrac{\delta J}{M} \partial_a f }
                        { \left| \partial_M f \right| + \dfrac{a}{M} \partial_a f }~.
    \end{equation}

    \item $\left[\partial_{M}f,\,\partial_{a}f,\,\partial_{Q}f\right] \equiv [-,-,+]$:
    
    \noindent Two subcases arise:
    \begin{enumerate}[label*=\arabic*.]
        \item For $\left| \partial_M f \right| > \dfrac{a}{M} \left| \partial_a f \right|$ and 
              $\delta Q\, \partial_Q f > \dfrac{\delta J}{M}\, \partial_a f$:
        \begin{equation}\label{--+i}
            \delta E < - \frac{ -\delta Q\, \partial_Q f + \dfrac{\delta J}{M} \left| \partial_a f \right| }
                            { \left| -\left| \partial_M f \right| + \dfrac{a}{M} \left| \partial_a f \right| \right| }~.
        \end{equation}

        \item For $\left| \partial_M f \right| > \dfrac{a}{M} \left| \partial_a f \right|$ and 
              $\delta Q\, \partial_Q f < \dfrac{\delta J}{M}\, \partial_a f$:
        \begin{equation}\label{--+ii}
            \delta E < \frac{ \left| -\delta Q\, \partial_Q f + \dfrac{\delta J}{M} \left| \partial_a f \right| \right| }
                            { \left| -\left| \partial_M f \right| + \dfrac{a}{M} \left| \partial_a f \right| \right| }~.
        \end{equation}
    \end{enumerate}

    \item $\left[\partial_{M}f,\,\partial_{a}f,\,\partial_{Q}f\right] \equiv [-,+,-]$:
    \begin{equation}\label{-+-}
        \delta E < \frac{ \left| \delta Q\, \partial_Q f - \dfrac{\delta J}{M} \partial_a f \right| }
                        { \left| \partial_M f \right| + \dfrac{a}{M} \partial_a f }, 
        \quad \text{with } \delta Q\, \partial_Q f - \dfrac{\delta J}{M} \partial_a f < 0~.
    \end{equation}

    \item $\left[\partial_{M}f,\,\partial_{a}f,\,\partial_{Q}f\right] \equiv [+,+,-]$:
    \begin{equation}\label{++-}
        \delta E < \frac{ \left| \delta Q\, \left| \partial_Q f \right| - \dfrac{\delta J}{M} \partial_a f \right| }
                        { \left| \partial_M f - \dfrac{a}{M} \partial_a f \right| },
    \end{equation}
    \vspace{-0.2cm}
    \begin{flushleft}
    with $\partial_M f < \dfrac{a}{M} \partial_a f$ and $\delta Q\, \left| \partial_Q f \right| > \dfrac{\delta J}{M} \partial_a f$.
    \end{flushleft}
\end{enumerate}

\bibliography{Overcharging.bib}
\bibliographystyle{utphys1}
 
 \end{document}